# Top Quark Mass Measurements

A.P. Heinson
(for the CDF and DØ Collaborations)

*Department of Physics and Astronomy*
*University of California, Riverside, CA 92521-0413, USA*

**Abstract.** First observed in 1995, the top quark is one of a pair of third-generation quarks in the Standard Model of particle physics. It has charge +2/3*e* and a mass of 171.4 GeV, about 40 times heavier than its partner, the bottom quark. The CDF and DØ collaborations have identified several hundred events containing the decays of top-antitop pairs in the large dataset collected at the Tevatron proton-antiproton collider over the last four years. They have used these events to measure the top quark's mass to nearly 1% precision and to study other top quark properties. The mass of the top quark is a fundamental parameter of the Standard Model, and knowledge of its value with small uncertainty allows us to predict properties of the as-yet-unobserved Higgs boson. This paper presents the status of the measurements of the top quark mass. It is based on a talk I gave at the Conference on the Intersections of Particle and Nuclear Physics in Puerto Rico, May 2006 [1], which also included discussion of measurements of other top quark properties.

## INTRODUCTION

### Top Quarks

The top quark is a spin half fermion with charge two thirds that of the positron. It is the heaviest of the six known quarks, and is the isospin partner of the bottom quark. Top quarks are produced in particle-antiparticle pairs at the Tevatron proton-antiproton collider at Fermi National Accelerator Laboratory in Batavia, Illinois. At the Tevatron's center-of-mass energy of 1.96 TeV, 85% of the production comes from quark-antiquark annihilation, and the remaining 15% from gluon-gluon fusion. The cross section for top quark pair production is $6.8 \pm 0.6$ pb [2,3], which is very small.

### The CDF and DØ Experiments

Two large multipurpose detectors sit on the Tevatron accelerator ring and measure the decay particles from the proton-antiproton collisions: CDF and DØ. These two collaborations observed the top quark in 1995 using roughly 50 $pb^{-1}$ of data [4,5]. Over ten years later, the Tevatron is still the only place where top quarks are produced and studied. The experiments have now collected datasets more than twenty times as large that they are using to make precision measurements of the top quark's properties.

## Top Quark Decay

It is not possible to detect top quarks directly, since they decay before they have time to hadronize. Almost 100% of the time, they decay to a *W* boson and a *b* quark. The *W* decays one ninth of the time to an electron and a neutrino, one ninth to a muon and neutrino, one ninth to a tau lepton and neutrino, and the remaining two thirds of the time to a pair of quarks. Since each top quark is produced with an antitop, there are thus two *b* quarks in the final state, which hadronize into jets of particles, and the particles from the decays of two *W* bosons. The top pair events are classified by the number of electrons and muons in the final state: zero = "alljets," one = "lepton+jets," and two = "dileptons." In order to identify top quark events from the overwhelmingly high background processes at a hadron collider, one must therefore reconstruct electrons, muons, light jets (from the *W* decay), *b* jets, and missing transverse energy (from the escaped neutrinos).

## Backgrounds

Most events that have the same particles in them as expected from top quark decay do not contain top quarks. After precise lepton identification has been applied, they are mostly events with a *W* or *Z* boson with some jets radiated from the initial-state quarks and antiquarks. The cross section for such processes is two or three orders of magnitude greater than that for top pair production. Other events that also look like signal include multijet events where one of the jets passes all the electron identification criteria, and bottom-antibottom pairs where one of the *b*'s decays to a muon that travels wide of the *b* jet such that it looks like it came from a *W* boson decay. For most $t\bar{t}$ decay channels, processes with a real *W* boson and real *b* jets are the most difficult to remove.

## *b*-Jet Identification

In order to reduce the *W*+jets background to only those ~1% of events containing *b* jets, CDF and DØ apply *b*-tagging algorithms to the jets in each event. These algorithms use information about the differences in properties between the decays of *b* hadrons and other ones. Fortunately, *b* mesons and baryons have a long lifetime before decaying to other particles, and since they are produced with large momentum, they travel several millimeters away from the primary interaction point before they decay. This distance can be measured using tracks reconstructed from high-precision vertex detectors made from silicon microstrip devices. The latest *b*-tagging algorithms train neural networks to recognize the difference between *b*-jets and charm-jets and light quark and gluon jets, and excellent separation is obtained for very low fake rates.

# TOP QUARK MASS

## Overview

The top quark's mass is a fundamental parameter of the Standard Model. For example, it enters into calculations of the *W* and *Z* boson masses through higher order loop effects. Of particular interest is that the coupling of the top quark to the Higgs boson (a force carrier for electroweak symmetry breaking and thus the generator of particle masses) is of order unity, very high, which forces the top quark mass to be large. The Higgs boson mass depends logarithmically on the top quark mass, and so a precise measurement of the top quark's mass leads to limits on the allowed mass of the Higgs boson. Figure 1 (a) shows how knowledge of the top quark mass has improved over the past 17 years, and Fig. 1 (b) shows how this now-precise measurement combined with the $W$ boson mass measurement restricts the allowed value of the Higgs boson mass in both the Standard Model and in the Minimal Supersymmetric Standard Model.

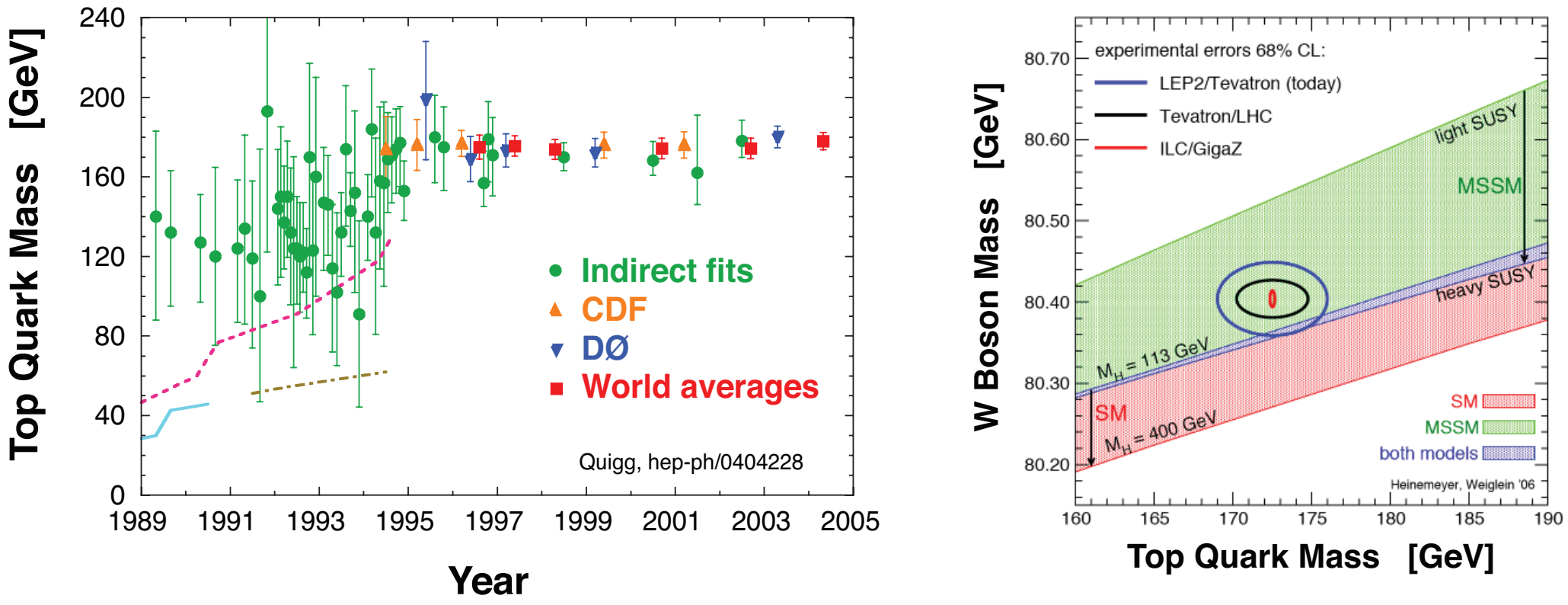

**FIGURE 1.** (a) A history of the top quark mass [6], showing excellent agreement between predictions from indirect fits to other data and direct measurements from the Tevatron experiments, and (b) limits on the Higgs boson mass [7] from measurements of the masses of the top quark and $W$ boson.

## DØ's Top Quark Mass Measurements

The DØ collaboration has developed several techniques for measuring the mass of the top quark over the past ten years. For dilepton events, there are two methods, the matrix-element-weighting method and the neutrino-weighting method [8], and for lepton+jets events, there is the matrix-element method [9]. They have also applied methods that use templates, and have adapted the Delphi experiment's ideogram method. Each of these techniques uses varying amounts of information from the signal and background events, and thus they have to some extent complementary systematic uncertainties. The matrix-element method is the most sensitive of these techniques. Improvements since publication two years ago include simultaneous fitting to the jet energy scale in order to calibrate it in situ, and use of the information of which jets are $b$-tagged. DØ's most precise measurement uses 370 pb$^{-1}$ of lepton+jets data to obtain $m_{\text{top}} = 170.6\ ^{+4.0}_{-4.7}\ \text{(stat)}\ \pm 1.4\ \text{(syst)}\ \text{GeV}$, illustrated in Fig. 2 (a). Figure 3 (a) shows this and other measurements of the top quark mass by DØ.

## CDF's Top Quark Mass Measurements

The CDF collaboration has also developed techniques for measuring the mass of the top quark, including a multivariate template method, a $b$-jet decay-length likelihood method, and the dynamic-likelihood method, which is a simplified version of DØ's matrix-element method. They have also applied a basic template method, the ideogram method, neutrino-weighting for dileptons, and the matrix-element method with in situ jet-energy scale calibration using *b*-tagged jet information. CDF's best result uses lepton+jets events from 940 $\mathrm{pb}^{-1}$ of data, and they find $m_{\mathrm{top}} = 170.9 \pm 2.2$ (stat) $\pm 1.4$ (syst) GeV. Figure 2 (b) shows this result, and Fig. 3 (b) shows a plot of this and other measurements by CDF [10–15].

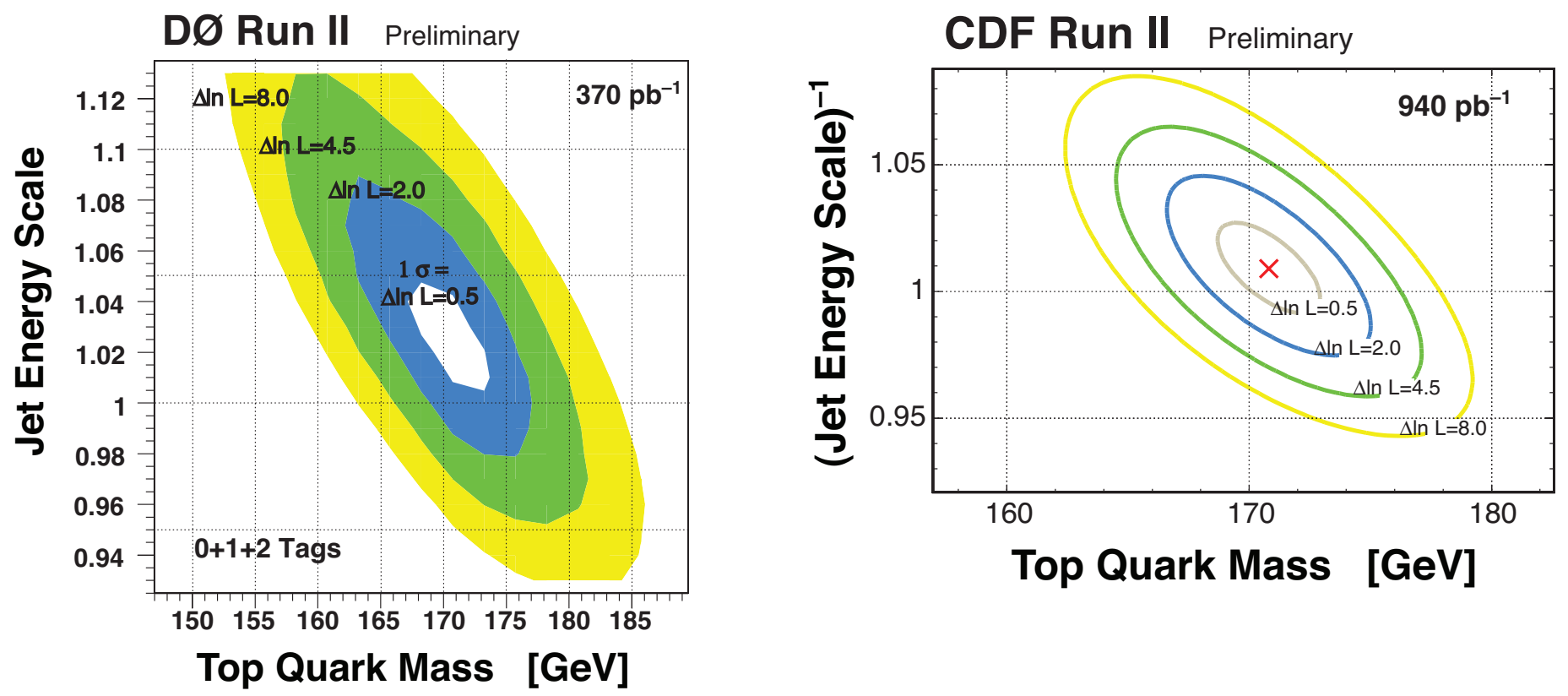

**FIGURE 2.** (a) DØ's measurement and (b) CDF's measurement of the top quark mass with in situ jet energy-scale calibration using the matrix-element method and lepton+jets events.

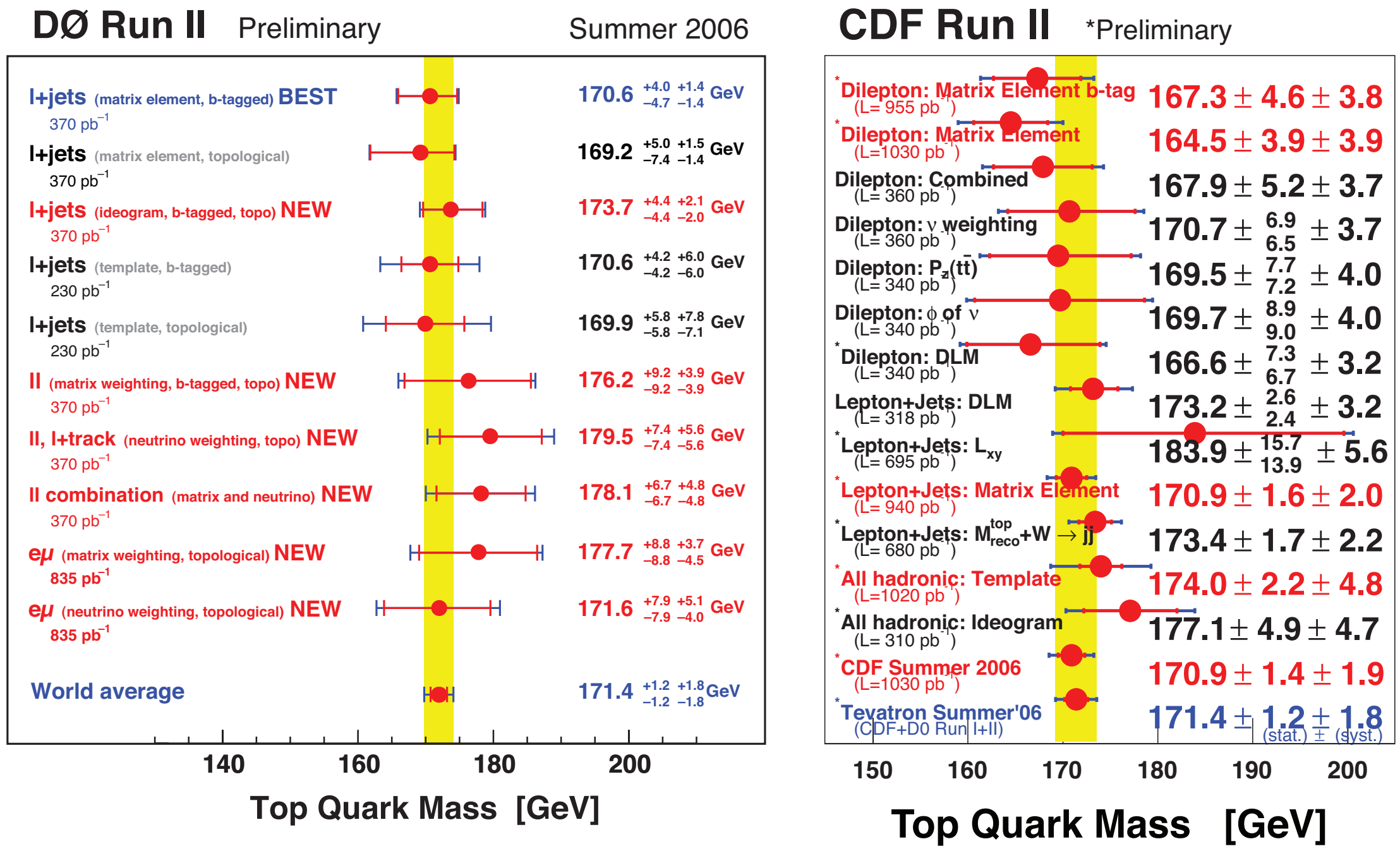

**FIGURE 3.** (a) DØ's measurements and (b) CDF's measurements of the top quark mass. DØ's most recent results use electron+muon events from 835 $\mathrm{pb}^{-1}$ of data and CDF's newest result uses dilepton events from 1030 $\mathrm{pb}^{-1}$ of data, first presented in July 2006.

## Combined Top Quark Mass Measurement

The top subgroup of the Tevatron Electroweak Working Group combines sets of top quark mass measurements to obtain a world average value, using the best measurements in orthogonal decay channels from each collaboration [16]. All correlated and uncorrelated components in the uncertainties are properly taken into account. The resulting latest value of the top quark mass is

$$m_{\mathrm{top}} = 171.4 \pm 1.2\ (\mathrm{stat}) \pm 1.8\ (\mathrm{syst})\ \mathrm{GeV}$$
$$= 171.4 \pm 2.1\ \mathrm{GeV},$$

which has a 1.2% uncertainty. As new measurements with larger datasets replace those used now, and as the systematic uncertainties become better understood, this uncertainty will continue to fall over the next few years. Figure 4 shows the measurements used to obtain this combined result.

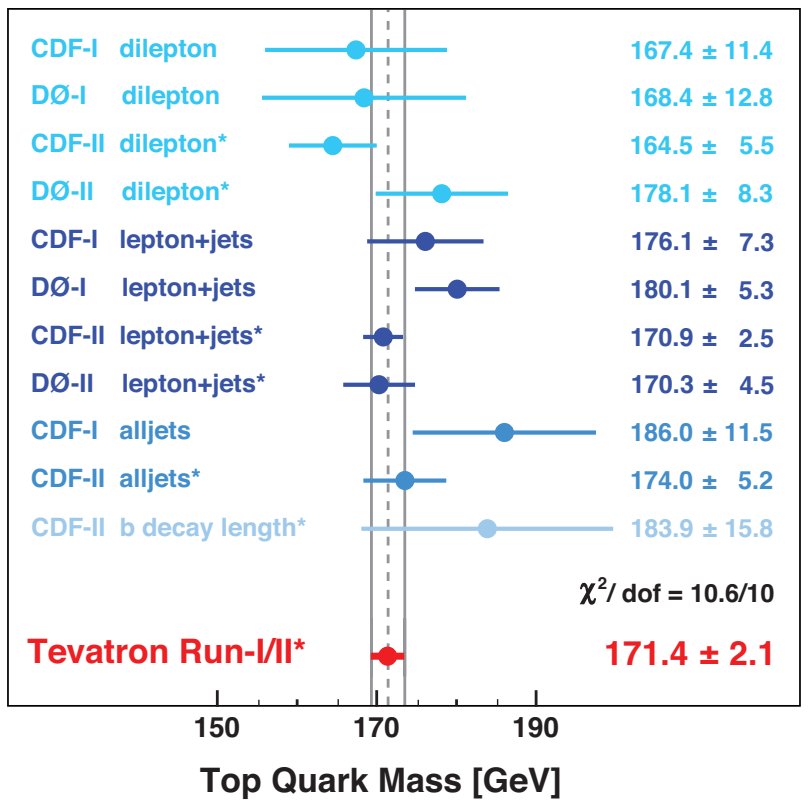

**FIGURE 4.** The mass of the top quark as calculated from eleven independent measurements made by the CDF and DØ collaborations [16]. A * indicates that a measurement has not been published yet.